\documentclass[11pt]{article}
\usepackage{graphics}
\newcommand{\Ds}{\!\not\!\!D}
\newcommand{\wt}[1]{\widetilde{#1}}
\newcommand{\Ca}{C_{\rm A}}
\newcommand{\Cf}{C_{\rm F}}
\newcommand{\Tf}{T_{\rm F}}
\newcommand{\Nl}{N_{\rm L}}
\newcommand{\ep}{\epsilon}
\newcommand{\om}{\omega}
\newcommand{\as}{\alpha_{\rm s}}
\newcommand{\mb}{\bar{\mu}}
\begin{document}
\thispagestyle{empty}
\setcounter{page}{0}
\begin{center}
\vspace*{\fill}
{\Large\bf Two-loop renormalization of the effective field theory\\[3pt]
of a static quark}
\vfill
{\large D.~J.~BROADHURST, A.~G.~GROZIN$^*$}
\vskip.7 cm
{\large Physics Department, \,Open University, \,Milton Keynes MK7 6AA,\,UK}
\end{center}
\vfill
{\bf Abstract }We give a recurrence relation for two-loop integrals
encountered in the effective field theory of an infinitely heavy quark,
$Q$, interacting with gluons and $\Nl$ massless quarks, $q$, from which
we obtain exact two-loop results, in any dimension and covariant gauge, for
the propagator of $Q$ and the vertex function of the heavy-light current
$J\equiv\overline{Q}\Gamma q$, at zero $q$\/-momentum. The anomalous
dimension of the $Q$\/-field agrees with the recent result of Broadhurst,
Gray and Schilcher. The anomalous dimension of the current is
\[\wt{\gamma}_{\rm J}\equiv\frac{{\rm d}
\log\wt{Z}_{\rm J}}{{\rm d}\log\mu}
=-\frac{\as}{\pi}\left\{1+
\left(\frac{127+56\zeta(2)-10\Nl}{72}\right)
\frac{\as}{\pi}
+O(\as^2)\right\}\]
which gives the new two-loop correction to the result of Voloshin
and Shifman.
\vfill
\hfill 15 March 1991

\hfill{OUT-4102-30}

{\footnotesize $^*$ permanent address: Institute of Nuclear Physics, 630090
Novosibirsk, USSR}

\newpage
\setlength{\parindent}{0.7cm}
\section{Introduction}

Recently an interesting new approach to QCD problems involving a heavy
quark was proposed, namely the effective field theory (EFT) of a static
quark~[1-5]. (See also the review~\cite{bjr}.) To leading order in the
$1/m$ expansion, the EFT lagrangian is~\cite{eah}
\begin{equation}
L=Q_0^{\dagger}{\rm i}D_0Q_0
+\overline{q}_0{\rm i}\Ds q_0
-\mbox{$\frac{1}{4}$}G_{0\mu\nu}^aG_{0\mu\nu}^a
-\left(\partial_{\mu}A_{0\mu}^a\right)^2\!\!/2a_0
+(\mbox{ghost term})
\label{lag}
\end{equation}
where $q_0$ and $A_{0\mu}$ are the bare light-quark and gluon fields of
conventional QCD and $Q_0$ is the bare static-quark field, which is a
two-component spinor. (When used as a 4-component spinor, its lower
components are assumed to vanish.) The free-field propagator of $Q_0$ is
$1/\om$ (equivalently, $(1+\gamma_0)/2\om$, in 4-component form) where
$\om\equiv p_0-m$ is the energy by which it is off-shell. Its coupling to
gluons is given by the vertex ${\rm i}g_0v_{\mu}t^a$, where $v_{\mu}
=(1,{\bf 0})$ is its velocity. The lagrangian~(\ref{lag}) possess an SU(2)
static-quark spin symmetry~\cite{wis}.

In this paper we consider the two-loop renormalization of EFT, using
dimensional regularization in $D\equiv4-2\ep$ dimensions. In the
$\overline{\rm MS}$ scheme the bare quantities in~(\ref{lag}) are related
to the corresponding renormalized quantities by
\[Q_0=\mb^{-\ep}\widetilde{Z}_2^{1/2}Q,\,\,
q_0=\mb^{-\ep}Z_2^{1/2}q,\,\,A_{0\mu}^a=\mb^{-\ep}Z_3^{1/2}A_{\mu}^a,\,\,
g_0=\mb^{\ep}Z_{\alpha}^{1/2}g,\,\, a_0=Z_3a\]
where $Z_2$, $Z_3$ and $Z_{\alpha}$ are the same as in QCD~\cite{nar}, with
$\Nl$ light-quark flavours, and $\mb^2=\mu^2{\rm e}^{\gamma}/4\pi$. Here we
calculate, to two loops, the static-quark wave-function renormalization
constant $\wt{Z}_2$ and the renormalization constant $\wt{Z}_{\rm J}$ of
the heavy-light current $J_0=\overline{Q}_0\Gamma q_0=\wt{Z}_{\rm J}J$,
where $\Gamma$ is some (irrelevant) gamma matrix. For $\wt{Z}_2$ we obtain
the same result as Broadhurst, Gray and Schilcher~\cite{mz2}, who have
independently extracted it from the gauge-invariant singularities of
on-shell wave-function renormalization of a finite-mass quark~\cite{mz1}.
>From $\wt{Z}_{\rm J}$ we obtain the two-loop correction to the
leading-order result~\cite{vas,paw} for the anomalous dimension
$\wt{\gamma}_{\rm J}\equiv{\rm d}\log\wt{Z}_{\rm J}/{\rm d}\log\mu$. This
is needed for the renormalization-group analysis of the extrapolations of
lattice computations of heavy-meson decay constants~\cite{lat}.

Renormalization of EFT corresponds closely to renormalization of the Wilson
line~[13-17], which coincides with the static-quark propagator in
coordinate space. In particular the lagrangian~(1) is essentially the same as
that proposed in~\cite{are}, for investigating Wilson lines, and used in the
two-loop calculations of~\cite{aoy,kr1}. In~\cite{aoy} Aoyama calculated
$\wt{Z}_2$ in the Feynman gauge, at $\Nl=0$; we disagree with his result.
In~\cite{kr1} Korchemsky and Radyushkin calculated the two-loop anomalous
dimension of a heavy-heavy current, which depends on the relative velocity
of initial and final quarks; the corresponding
one-loop result~\cite{pol} has recently been rediscovered in EFT~\cite{fgg}.

\section{One- and two-loop integrals}

The static-quark bare self-energy term $-{\rm i}\Sigma(\om)$ is given, to
two loops, by the diagrams of Fig~1, where the blob in Fig~1a includes the
light-quark, gluon and ghost loops in the gluon propagator. We take the
light quarks as massless and easily evaluate the master one-loop EFT
integral
\[\int{\rm d}^D k
\left(\frac{-1}{k^2}\right)^{\alpha}
\left(\frac{\om}{\om+k_0}\right)^p
\equiv{\rm i}\pi^{D/2}(-2\om)^{(D-2\alpha)}I(\alpha,p)\]
using the parameterization
\[\frac{1}{A^{\alpha}P^p}=\frac{\Gamma(\alpha+p)}{\Gamma(\alpha)\Gamma(p)}
\int_{0}^{\infty}\frac{y^{p-1}{\rm d}y}{(A+y P)^{\alpha+p}}\]
where we adopt the convention that massless lines have greek indices and
static lines have roman ones. We find that
\begin{equation}
I(\alpha,p)=\frac{\Gamma(2\alpha+p-D)\Gamma(D/2-\alpha)}
{\Gamma(\alpha)\Gamma(p)}
\label{I2}
\end{equation}
which enables us to evaluate diagrams of the form of Fig~1a.

The master two-loop EFT integral, for diagrams of the form of Fig~1b, is
\[\int{\rm d}^D k\int{\rm d}^D l
\left(\frac{-1}{k^2}\right)^{\alpha}
\left(\frac{-1}{l^2}\right)^{\beta}
\left(\frac{-1}{(k-l)^2}\right)^{\gamma}
\left(\frac{\om}{\om+k_0}\right)^p
\left(\frac{\om}{\om+l_0}\right)^q
\]\[
\equiv-\pi^{D}(-2\om)^{2(D-\alpha-\beta-\gamma)}I(\alpha,\beta,\gamma,p,q)\]
In degenerate cases one uses~(\ref{I2}) and the corresponding
result~\cite{ibp} for the master one-loop diagram of massless QCD
\[G(\alpha,\beta)=\frac{\Gamma(\alpha+\beta-D/2)\Gamma(D/2-\alpha)
\Gamma(D/2-\beta)}{\Gamma(\alpha)\Gamma(\beta)\Gamma(D-\alpha-\beta)}\]
to obtain
\begin{eqnarray}
I(\alpha,\beta,0,p,q)&=&I(\alpha,p)I(\beta,q)
\label{dg1}\\
I(\alpha,0,\gamma,p,q)&=&I(\alpha,2\gamma+p+q-D)I(\gamma,q)
\label{dg2}\\
I(\alpha,\beta,\gamma,p,0)&=&I(\alpha+\beta+\gamma-D/2,p)
\label{dg3}
G(\beta,\gamma)
\end{eqnarray}
which suffice for diagrams of the form of Fig~1c and Fig~1d, since the
integrand of Fig~1c has only 4 factors in the denominator and the integrand
of Fig~1d can be expressed as the sum of three such terms after multiplying
by
\[1=\frac{1}{\om}\left\{(\om+k_0)+(\om+l_0)-(\om+k_0+l_0)\right\}\]
This is an enormous simplification in comparison with on-shell massive QCD
calculations~\cite{mz2,mz1}, where this last diagram was the most
difficult.

In general we use a recurrence relation to reduce the non-degenerate case
of the master integral to a sum of known degenerate cases. Using the method
of integration by parts~\cite{mz1,ibp}, we have found that
\begin{equation}
(D-\alpha-2\gamma-p-q+1)I=\left\{\alpha{\bf A}^+({\bf \Gamma}^--{\bf B}^-)
+(2(D-\alpha-\beta-\gamma)-p-q+1){\bf Q}^-\right\}I
\label{rec}
\end{equation}
where, for example, ${\bf A}^+$ raises $\alpha$ and ${\bf Q}^-$ lowers $q$.
Again this is simpler than the triangle relation~\cite{ibp} of massless
QCD, because no more than three terms appear and no static-line index is
increased.

Now it is a matter of programming to reduce every one-loop integral to a
rational function of $D\equiv4-2\ep$ times
\begin{equation}
\Gamma_1\equiv(-2\om)^{-2\ep}\Gamma(-\ep)\Gamma(1+2\ep)/(4\pi)^{D/2}
\label{G1}
\end{equation}
and every two-loop integral from Fig~1 to a linear combination of
$\Gamma_1^2$ and
\begin{eqnarray}
\Gamma_2&\equiv&(-2\om)^{-4\ep}\Gamma^2(-\ep)\Gamma(1+4\ep)/(4\pi)^{D}
\label{G2}\\
&=&(1+4\zeta(2)\ep^2-16\zeta(3)\ep^3)\Gamma_1^2+0(\ep^2)
\label{zeta}
\end{eqnarray}
with coefficients which are rational functions of $D$. We used
REDUCE~\cite{red} on a Vax and checked results with Mathematica~\cite{mma}
on a Macintosh. The programming effort and run times were far less than in
the related QCD calculations~\cite{mz2,mz1}.

\section{Anomalous dimension of the static-quark field}

Using these methods, it is straightforward to compute the two-loop
corrections to the bare propagator $S_0(\om)$. We find that
\begin{eqnarray}
\om S_0(\om)&=&\frac{\om}{\om-\Sigma(\om)}=1+Z_1\Gamma_1\Cf
g_0^2+\left\{Z_{{\rm A}1}\Ca\Gamma_1^2\right.
\nonumber\\
&&\left.\phantom{\Gamma_1^2}+
(Z_{{\rm A}2}\Ca+Z_{{\rm F}2}\Cf+Z_{{\rm L}2}\Tf\Nl)
\Gamma_2\right\}\Cf g_0^4+O(g_0^6)
\label{Zw}
\end{eqnarray}
with coefficients
\begin{eqnarray*}
Z_1&=&a_0-\frac{D-1}{D-3}\\
Z_{{\rm A}1}&=&-{Z_1\over D-3}\\
Z_{{\rm A}2}&=&{(D-2)^2(D-5)\over2(2D-7)(D-3)^3(D-6)}
+{(D^2-4D+5)Z_1\over2(D-3)^2(D-6)}
-{(D^2-9D+16)Z_1^2\over8(D-3)(D-6)}\\
Z_{{\rm F}2}&=&{Z_1^2\over2}\\
Z_{{\rm L}2}&=&{2(D-2)\over(2D-7)(D-3)(D-6)}
\end{eqnarray*}
and colour factors $\Ca=N$, $\Cf=(N^2-1)/2N$, $\Tf=\frac{1}{2}$, for an
SU($N$) gauge group, or $\Ca=0$, $\Cf=\Tf=1$ for a U(1) gauge group.

We now renormalize the bare static-quark field $Q_0=\mb^{-\ep}
\wt{Z}_2^{1/2}Q$ using the minimal wave-function renormalization constant
$\wt{Z}_2$ which makes the renormalized propagator $S(\om)=S_0(\om)/
\wt{Z}_2$ finite. (Tildes are used here to distinguish EFT renormalization
constants from those of QCD.) We work in the $\overline{\rm MS}$ scheme, in
which the bare coupling and gauge parameter are renormalized by~\cite{nar}
\begin{eqnarray}
\frac{g_0^2}{g^2\mb^{2\ep}}&=&Z_{\alpha}=
1-\frac{\as}{4\pi\ep}\left(\frac{11}{3}C_{\rm A}-\frac{4}{3}T_{\rm F}N_{\rm
L}\right)+O(\as^2)
\label{as}\\
\frac{a_0}{a}&=&Z_3=
1+\frac{\as}{4\pi\ep}
\left(\frac{13-3a}{6}C_{\rm A}-\frac{4}{3}T_{\rm F}N_{\rm
L}\right)+O(\as^2)
\label{ab}
\end{eqnarray}
where $\as\equiv g^2/4\pi$. We posit a minimal form $\wt{Z}_2=
1+(C_{11}(a)/\ep)\as/\pi+(C_{22}(a)/\ep^2+C_{21}(a)/\ep)\as^2/\pi^2+O(\as^3)$
and determine its coefficients from~(\ref{Zw}). The renormalizability of
the theory requires that this constant make $S_0(\om)/\wt{Z}_2$ finite for
all $\om$, which provides a highly non-trivial check on our result. Then,
using the $\mu$ dependence of $\as$ and $a$, implied
by~(\ref{as},\ref{ab}), we obtain the anomalous dimension $\wt{\gamma}_{\rm
F}\equiv{\rm d}\log\wt{Z}_2/{\rm d}\log\mu$ of the static $Q$ field. As a
check on the whole procedure we mirror it, step by step, by that for a
massless quark, to obtain the corresponding QCD quantity $\gamma_{\rm F}$
for the massless $q$ field from the recurrence relations of~\cite{ibp}. Our
result is
\begin{eqnarray}
\wt{\gamma}_{\rm F}&=&\frac{(a-3)\Cf\as}{2\pi}
+\left\{\left(\frac{a^2}{32}
+\frac{a}{4}
-\frac{179}{96}\right)\Ca
+\frac{2}{3}\Tf\Nl\right\}
\frac{\Cf\as^2}{\pi^2}
+O(\as^3)
\nonumber\\
&=&\gamma_{\rm F}-\frac{3\Cf\as}{2\pi}
-\left\{\frac{127}{48}\Ca
-\frac{3}{16}\Cf
-\frac{11}{12}\Tf\Nl\right\}
\frac{\Cf\as^2}{\pi^2}
+O(\as^3)
\label{gfd}
\end{eqnarray}
where the latter form was independently obtained in~\cite{mz2} from the
gauge-invariant singularities of {\em on\/}-shell wave-function
renormalization of a {\em finite\/}-mass quark. Note that~(\ref{gfd})
vanishes in QED in the renormalized Yennie gauge~\cite{a03} $a=3$, for
which there is no `infrared catastrophe'~\cite{irc}.
From~(\ref{gfd}) we find that $\gamma_{\rm F}$ agrees with the
QCD result of Egorian and Tarasov~\cite{eat} (misquoted in~\cite{nar},
which contains sign errors).

Setting $a=1$ and $\Nl=0$ in~(\ref{gfd}), we
discover an error in~\cite{aoy}, where the renormalization of the Wilson
line was calculated (neglecting light-quark loops) in the Feynman gauge
(where Fig~1b gives no contribution). The $\as^2/\ep^2$ terms in~\cite{aoy}
are inconsistent with the renormalization group and the $\as^2/\ep$ terms
include the contributions of Figs~1c,d with the wrong sign. This incorrect
result was used in~\cite{kr2}, where the authors also omitted a factor of 2
from the two-loop term in the anomalous dimension. In view of these
approximations and errors the two-loop analysis of~\cite{kr2}
needs reconsideration. 

\section{Anomalous dimension of the heavy-light current}

The simplest method for finding the anomalous dimension $\wt{\gamma}_{\rm
J}$ of the current $J_0=\overline{Q}_0\Gamma q_0$ is to consider the bare
vertex function with zero light-quark momentum. The diagrams of Fig~2 then
differ from those of Fig~1 only in their numerator structure and the
indices of their denominators. The effect is to multiply the gamma matrix
$\Gamma$ by a factor $\Gamma_0(\om)$, which is gauge-dependent and more
complicated than~(\ref{Zw}). But it becomes much simpler when one
multiplies by the external static-quark bare propagator $S_0(\om)$. We
find that
\begin{eqnarray}
V_0(\om)&\equiv&\om S_0(\om)\Gamma_0(\om)=1+V_1\Gamma_1\Cf
g_0^2+\left\{(V_{{\rm A}1}\Ca+V_{{\rm F}1}\Cf)\Gamma_1^2\right.
\nonumber\\
&&\left.\phantom{\Gamma_1^2}+
(V_{{\rm A}2}\Ca+V_{{\rm F}2}\Cf+V_{{\rm L}2}\Tf\Nl)
\Gamma_2\right\}\Cf g_0^4+O(g_0^6)
\label{Vw}
\end{eqnarray}
is gauge-invariant. The coefficients are
\begin{eqnarray*}
V_{1} &=&-{D-1\over D-3}\\
V_{{\rm A}1}&=&{D^2-4D-1\over2(D-3)^2(D-4)}\\
V_{{\rm F}1}&=&{D-2\over(D-3)(D-4)}\\
V_{{\rm A}2}&=&-{2D^4-22D^3+89D^2-151D+78\over2(2D-7)(D-3)^2(D-4)(D-6)}\\
V_{{\rm F}2}&=&-{(D^2-5D+2)D\over2(D-3)^2(D-4)(D-6)}\\
V_{{\rm L}2}&=&{2(D-2)\over(2D-7)(D-3)(D-6)}
\end{eqnarray*}
independently of the spin structure of the current $J$.

The renormalization constant $\wt{Z}_{\rm J}=\wt{Z}_{\Gamma}\wt{Z}_2^{1/2}
Z_2^{1/2}$ may be calculated from the requirement that the renormalized
vertex function $\Gamma(\om)=\Gamma_0(\om)/\wt{Z}_{\Gamma}$ be finite.
Equivalently, but more conveniently, we have $\wt{Z}_{\rm
J}=\wt{Z}_{V}\wt{Z}_2^{-1/2}Z_2^{1/2}$, where $\wt{Z}_{\rm
V}=\wt{Z}_{\Gamma}\wt{Z}_2$ makes $V(\om)=V_0(\om)/\wt{Z}_{\rm V}$ finite.
This gives two highly non-trivial checks on our results. First, the
$\as^2/\ep^2$ terms in~(\ref{Vw}) must be such that division by a constant
makes $V(\om)$ finite for all $\om$. Second, the anomalous dimension
$\wt{\gamma}_{\rm V}={\rm d}\log\wt{Z}_{\rm V}/{\rm d}\log\mu$ must be
gauge-invariant, to ensure that $\wt{\gamma}_{\rm J}=\wt{\gamma}_{\rm V}
-\frac{1}{2}\wt{\gamma}_{\rm F} +\frac{1}{2}{\gamma}_{\rm F}$ is also
gauge-invariant. As a third check, we mirrored this EFT calculation, step
by step, by the QCD calculation of $\gamma_{\overline{q}q}$, obtaining
agreement with Tarrach~\cite{tar}. These checks give us considerable
confidence in our final result
\begin{eqnarray}
\wt{\gamma}_{\rm J}&=&-\frac{3\as\Cf}{4\pi}
-\left\{\frac{49}{96}\Ca-\frac{5}{32}\Cf-\frac{5}{24}\Tf\Nl
-(\mbox{$\frac{1}{4}$}\Ca-\Cf)\zeta(2)\right\}
\frac{\Cf\as^2}{\pi^2}+O(\as^3)
\nonumber\\
&=&\mbox{$\frac{1}{2}$}\gamma_{\overline{q}q}
+\left\{
\mbox{$\frac{1}{2}$}\Ca
+\mbox{$\frac{1}{4}$}\Cf
+\left(\mbox{$\frac{1}{4}$}\Ca-\Cf\right)\zeta(2)\right\}
\frac{\Cf\as^2}{\pi^2}+O(\as^3)
\label{gjd}
\end{eqnarray}
whose second form we have also verified from the quark-condensate
contributions to the correlator of $J$~\cite{ef2}.

\section{Summary and conclusions}

We have given a method for calculating one-scale one- and two-loop EFT
integrals. The master two-loop integral can be reduced by the recurrence
relation~(\ref{rec}) to the degenerate
cases~(\ref{dg1},\ref{dg2},\ref{dg3}), which yield the
structures~(\ref{G1},\ref{G2}). We have used this method to calculate the
static-quark propagator and a particular case of the heavy-light vertex
function, obtaining~(\ref{Zw},\ref{Vw}) in an arbitrary spacetime dimension
and in any covariant gauge. From these follow the anomalous dimensions of
the static-quark propagator~(\ref{gfd}) and the heavy-light
current~(\ref{gjd}). The first agrees with~\cite{mz2} and contradicts an
old result for the Wilson line~\cite{aoy}. The second is also confirmed
in~\cite{ef2}. It gives the radiative correction to the one-loop result
of~\cite{vas,paw} and is necessary for matching QCD and EFT currents, with
account of the radiative corrections of~\cite{eah}. The lack of it
in~\cite{lat} limited the accuracy of extrapolation of lattice computation.

Novel features of EFT calculations include: the trivial simplification of
diagrams such as Figs~1d and~2i, which are the most difficult in on-shell
massive QCD calculations~\cite{mz2,mz1}; an algorithm more efficient than
the triangle relation~\cite{ibp} of massless QCD; the appearance of
$\zeta(2)=\pi^2/6$ in counterterms, via the relation~(\ref{zeta}) between
the basic integral structures, in contrast to the situation in QCD, where
$\zeta(3)$ is the first term to appear in the expansion.

{\bf Acknowledgement} We are most grateful to SERC for a grant which made
our collaboration possible.

\raggedright
\setlength{\parindent}{0cm}

\newpage
\begin{figure}[h]
\centerline{\includegraphics{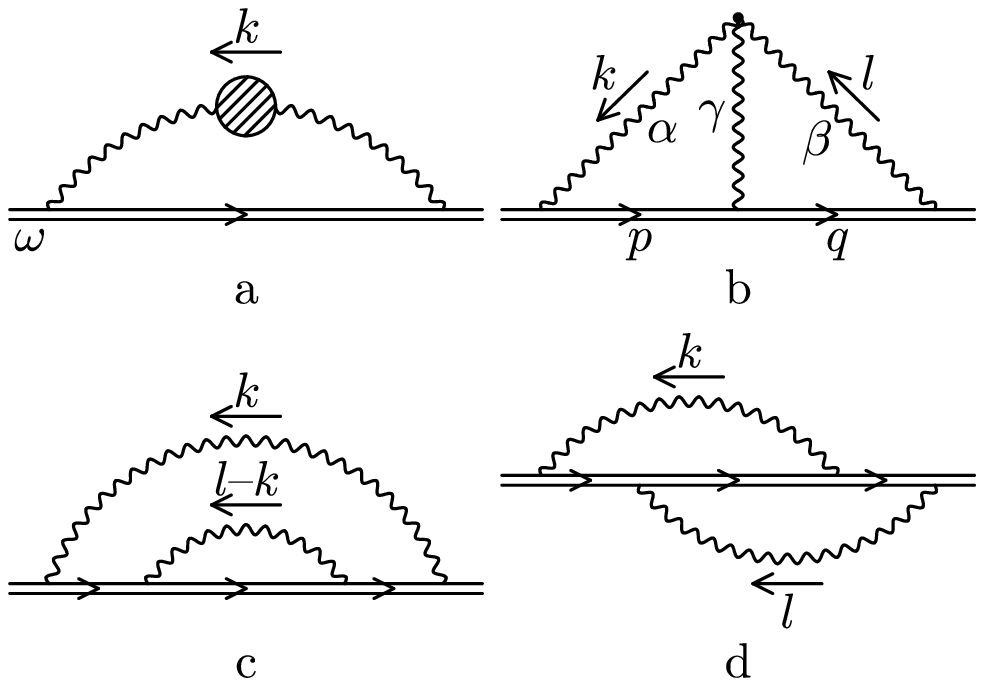}}
\caption{Heavy-quark self-energy diagrams, to two loops}
\end{figure}
\begin{figure}[h]
\centerline{\includegraphics{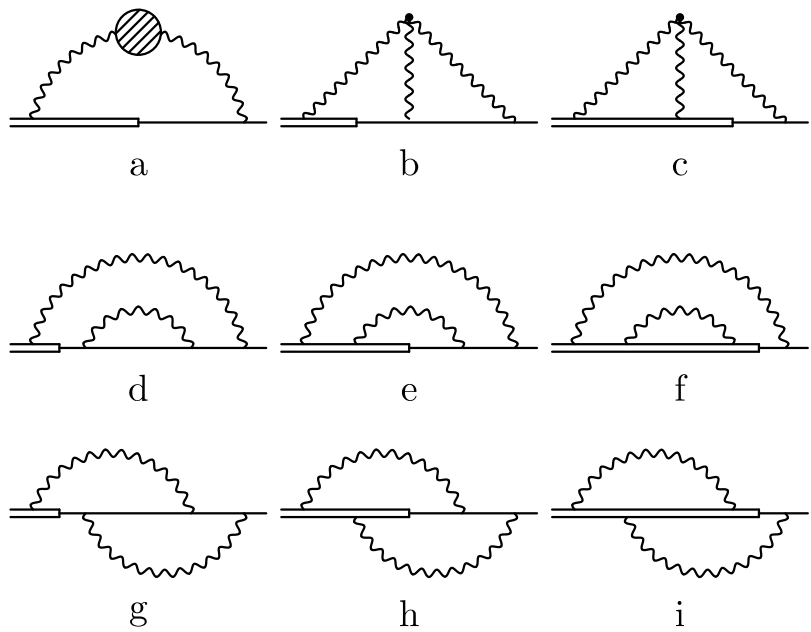}}
\caption{Heavy-light proper vertex diagrams, to two loops}
\end{figure}
\thispagestyle{empty}
\end{document}